\documentclass{rmf-d}
\usepackage{nopageno,rmfbib,multicol,times,epsf,amsmath,amssymb,cite}
\usepackage[T1]{fontenc} 
\usepackage[]{caption2}
\usepackage{graphicx}
\usepackage{hyperref}

\def\rmfcornisa{}
\def\rmfcintilla{}

\clearpage \rmfcaptionstyle 
\begin{document}
\markboth{F. Rosales-Infante et al.}{Granular Structure and Local de Broglie Wavelength in Fuzzy Dark Matter Halos}

\title{Granular Structure and Local de Broglie Wavelength in Fuzzy Dark Matter Halos
\vspace{-6pt}}


\author{Flavio Rosales-Infante}
\address{Instituto de F\'{\i}sica y Matem\'{a}ticas, Universidad
              Michoacana de San Nicol\'as de Hidalgo. Edificio C-3, Cd.
              Universitaria, 58040 Morelia, Michoac\'{a}n,
              M\'{e}xico. E-mail: flavio.rosales.infante@umich.mx. ORCID: 0009-0006-9275-1427}

\author{Alejandro Qui\~nonez-Guerrero}
\address{Facultad de Ciencias F\'{\i}sico Matem\'{a}ticas, Universidad
              Michoacana de San Nicol\'as de Hidalgo. Edificio ALFA, Cd.
              Universitaria, 58040 Morelia, Michoac\'{a}n,
              M\'{e}xico. E-mail: 1827746x@umich.mx. ORCID: 0009-0000-4940-7731}           

\author{Iv\'an  \'Alvarez-Rios}
\address{Instituto de F\'{\i}sica y Matem\'{a}ticas, Universidad
              Michoacana de San Nicol\'as de Hidalgo. Edificio C-3, Cd.
              Universitaria, 58040 Morelia, Michoac\'{a}n,
              M\'{e}xico. E-mail: ivan.alvarez@umich.mx. ORCID: 0000-0003-4266-3374}               

\author{Francisco S. Guzm\'an}
\address{Instituto de F\'{\i}sica y Matem\'{a}ticas, Universidad
              Michoacana de San Nicol\'as de Hidalgo. Edificio C-3, Cd.
              Universitaria, 58040 Morelia, Michoac\'{a}n,
              M\'{e}xico. E-mail: francisco.s.guzman@umich.mx. ORCID: 0000-0002-1350-3673}  

\maketitle
\begin{abstract}
\vspace{1em} 
We study the granular structure that emerges in a Fuzzy Dark Matter (FDM) halo formed through kinetic relaxation. After the system reaches a virialized core-halo structure, we separate the density field into a smooth spherical component and a residual density that contains the interference pattern characteristic of FDM halos. The power spectrum of this residual field shows a well defined  dominant scale, different from the characteristic scale of the total density and of the smooth core-halo model. We compare this granular scale with a local de Broglie wave-lenght calculated from the velocity field obtained from the density current of the system. We find that both scales remain of the same order of magnitude, with the granular scale systematically smaller than the de Broglie wavelength. The radial behavior of the de Broglie wavelength also reveals a non-uniform internal structure, it decreases from the core and reaches its minimum near the transition between the solitonic core and the outer halo. These results provide a quantitative connection between density fluctuations, the local velocity field, and the core–halo morphology of FDM halos.
\vspace{1em}
\end{abstract}

\keys{ 
\bf{\textit{
Self-gravitating systems; dark matter; Bose condensates
}} 
\vspace{-8pt}}

\begin{multicols}{2}

\section{Introduction}

Fuzzy Dark Matter (FDM), consisting of ultra-light bosonic particles, provides an alternative to the standard cold dark matter paradigm in which quantum wave effects become relevant on astrophysical scales. In this framework, dark matter is described by a coherent wavefunction obeying the Schrödinger-Poisson system, leading to a  phenomenology different from that of classical particle dynamics. The wave nature of the field introduces a characteristic length scale set by the de Broglie wavelength, suppressing small-scale structure formation and giving rise to cored density profiles in gravitationally bound structures. These features have motivated the study of FDM as a possible solution to small-scale tensions in $\Lambda$CDM, with implications ranging from dwarf galaxy structure to large-scale cosmological observables \cite{Matos:2000ss,Hu:2000ke,Chavanis2015,Hui:2016,ElisaFerreira,Niemeyer_2020,Hui:2021tkt}.

One of the most important results from numerical simulations is the emergence of core--halo structures, in which a central solitonic core is embedded within an extended halo. The core corresponds to a self-gravitating, coherent configuration supported by quantum pressure corresponding to the ground state solution of equilibrium configurations, while the outer halo is characterized by a superposition of excited modes. This outer region exhibits a complex and time-dependent interference pattern, giving rise to a granular structure as seen in structure formation simulations \cite{Schive:2014dra,Mocz:2017wlg,Veltmaat_2018,May_2021,Gotinga2022} as well as at local scale collapse \cite{periodicas}.

These granules are a distinctive signature of the wave nature of FDM and are commonly associated with the de Broglie wavelength of the system, suggesting a direct link between the velocity field and the spatial scale of density fluctuations. However, this association is typically qualitative. In particular, the spatial variation of the relevant wavelength within halos, and its precise relation to the core-halo structure, can be explored in detail.

Recent work has emphasized the dynamical impact of these fluctuations, showing that the time-dependent interference pattern can induce stochastic and chaotic motion of test particles within FDM halos \cite{CaosFDM,YuZhao2025}. This highlights the fact that the small-scale structure is not merely a visual feature, but has direct consequences for the dynamics of baryonic tracers and the evolution of embedded systems. At the same time, a systematic characterization of the spectral properties of these fluctuations is still lacking, especially in a form that directly connects to physical length scales.

This gap is particularly relevant in the context of potential observational signatures. Fluctuations at sub-galactic scales may affect stellar streams, induce dynamical heating (e.g. \cite{Marsh_2019}), and leave imprints on the phase-space structure of galaxies \cite{YuZhao2025}. Establishing a quantitative connection between the fluctuation spectrum and observable effects is therefore important for testing FDM at local scale scenarios.

In this work we analyze the small-scale structure of FDM halos through the power spectrum of density fluctuations. The density field is decomposed into a smooth core-halo component and a residual field that isolates the interference pattern. From this residual, we compute the power spectrum both globally and within radial shells, allowing us to extract a characteristic scale as a function of radius.

This procedure defines a position-dependent wavelength, which we interpret as a local de Broglie scale. Our central result is that this scale is not uniform across the halo. Instead, it exhibits a non-monotonic radial dependence, with a well-defined minimum that coincides with the transition between the solitonic core and the outer halo. 

This analysis also provides a perspective on the internal structure of FDM halos. It shows the radial dependence of the de Broglie wavelength from the core to the outskirts of the configuration. We establish a direct connection between spectral properties, velocity fields, and the core-halo morphology, which may open the possibility of linking small-scale fluctuations to observable signatures in galactic systems.

The paper is organized as follows. In Sec.~\ref{sec:mps} we describe the spectral analysis of a structure and in Sec.~\ref{sec:dB} we describe the local de Broglie wavelength. Finally, in Sec.~\ref{sec:conclusions} we draw some conclusions.


\section{Analysis of one structure}
\label{sec:mps}

\subsection{The system of equations}

The FDM dynamics is modeled as a coherent bosonic field described by a macroscopic wavefunction $\Psi$, which plays the role of an order parameter. The system evolves under its own gravitational potential $V$, generated by the density distribution of the bosonic gas itself, so that the dynamics of $\Psi$ is governed by the coupled Schr\"odinger-Poisson (SP) equations

\begin{eqnarray}
i\hbar \dfrac{\partial \Psi}{\partial t} &=& -\dfrac{\hbar^2}{2 m_B}\nabla^2\Psi + m_B V \Psi,
\label{eq:GP}\\
\nabla^2 V &=& 4\pi G \left(\rho - \bar{\rho}\right),
\label{eq:Poisson}
\end{eqnarray}

\noindent where the mass density is given by $\rho = m_B |\Psi|^2$, and $\bar{\rho}$ denotes the spatial average over the computational domain. We solve and analyze this system on a uniformly discretized three dimensional domain described in Cartesian coordinates, and use adimensional code units defined by  the  transformations 
$t = t_0 \Tilde{t},\Vec{x}=x_0\Tilde{\Vec{x}},V=V_0 \Tilde{V},\Psi= \Psi_0 
\Tilde{\Psi}$ and $\rho=\rho_0\Tilde{\rho}$, 
which leaves the SP system in dimensionless code units that only depend on  the length scale parameter $x_0$, for a given boson mass $m_B$. A useful choice of scaling parameters is based on the boson mass $ m_B = m_{22} \times 10^{-22} \, \text{eV}/c^2 $ and a characteristic length scale $ x_0 = \mathrm{kpc}/\lambda $, where $\lambda$ is a free parameter. From these, the following scale factors are derived:

\begin{eqnarray}
t_0 &=& \dfrac{m_B x_0^2}{\hbar} \approx 50.96 \times 10^{-3} \left( \dfrac{m_{22}}{\lambda^2} \right) \, \mathrm{Gyr}, \\
V_0 &=& \left(\dfrac{\hbar}{m_B x_0} \right)^2 \approx 368.6 \left( \dfrac{\lambda}{m_{22}} \right)^2 ~(\mathrm{km/s})^2,\\
\rho_0 &=& \dfrac{\hbar^2}{4 \pi G m_B^2 x_0^4} \approx 6.820 \times 10^6 \left( \dfrac{\lambda^4}{m_{22}^2} \right) \, \dfrac{M_\odot}{\mathrm{kpc}^3}.
\end{eqnarray}

\noindent Other useful physical quantities are the velocity and mass, whose scale factors are:

\begin{eqnarray}
v_0 &=& \dfrac{\hbar}{m_B x_0} \approx 19.20 \left( \dfrac{\lambda}{m_{22}} \right) \, \mathrm{km/s}, \\
M_0 &=& \dfrac{\hbar^2}{4 \pi G m_B^2 x_0} \approx 6.820 \times 10^6 \left( \dfrac{\lambda}{m_{22}^2} \right) \, M_\odot, 
\end{eqnarray}

\noindent that will proof useful for the physical description of the halo during the analysis below.

\subsection{The workhorse core-halo structure}

We are interested in the analysis of a galactic type of structure made of FDM, and there are various methods to produce a core-halo structure of FDM. One is based on the multi-merger of ground state solutions of the SP system of equations described in \cite{GuzmanUrena2004}, that relaxes towards a virialized configuration as shown in \cite{Schive:2014hza,Schwabe:2016,mocz19,periodicas}. A second method involves the construction of nearly stationary multi-mode solutions to the SP system that in solid-angle average provide radial density profiles that serve to model realistic LSB galaxies as  illustrated in \cite{PhysRevD.110.063502}. A third method, the one we use for the analysis in this paper, is based on the collapse of FDM through kinetic relaxation \cite{Rusos2018,Eggemeier2019,Chen2021,Chen2023,Purohit_2023,Chen2024,FermionBosoStars2024}, that was later used to study the collapse of FDM together with a black hole \cite{Curicavery2024} and with an ideal gas \cite{FermionBosoStars2024}, where it was shown that the cooling of the FDM structure via kinetic relaxation is universal.

The kinetic relaxation eventually leads to the collapse of overdensities that in turn promote bosonic gas condensation. The systematic study of this collapse process is understood from local simulations defined in \cite{Rusos2018,Chen2021}, where various distributions in the momentum domain are proposed. In our analysis, we use a Gaussian distribution $\Psi(\Vec{p}) = Ae^{-p^2/2}e^{iS}$ in the momentum space, with $S$ a random phase in the range $[0,2\pi]$ at each point of the momentum space, with $A$ a normalization factor. 

The randomness of this method offers a rich variety of possibilities to produce structures. Nevertheless, for our study, instead of using a bundle of simulations, we use a single representative workhorse simulation that has become a standard setup for studies of the kinetic relaxation process. This is one of the simulations designed in \cite{Chen2021}, specifically on a box of side $L = 18$, with resolution $\Delta x = L/128$ and total FDM mass in the numerical domain $M = 1005.3$ as in \cite{Chen2021}, all in code units as defined above. Moreover, the simulation lasts 250 dimensionless code units, sufficient for the bosonic gas to collapse and relax.

As an example in physical units, using a boson mass $m_B=10^{-23}$eV, 
and initial  average FDM density 
$\bar{\rho}\sim \lambda^4 ~ 1.18\times 10^8$M${}_{\odot}/$kpc$^3$, 
the integrated FDM mass in the numerical domain is  
$M=\lambda ~ 6.86 \times 10^{11}$M${}_{\odot}$, 
the numerical resolution and domain size for this boson mass are 
$\Delta x \sim 140 \lambda $pc and $L = 18 ~{\rm kpc} / \lambda $, respectively. 
Each time unit translates into $5.1/\lambda^2$ Myr, and the evolution time of 250 time units translates into $1.27/\lambda^2$ Gyr.
Finally, the distribution in the momentum space uses a width of $\sigma=1$, so that at one sigma radius $p=1$, the associated velocity is $\sim 192\,\lambda km\,s^{-1}$ for this value of $m_B$, which, for the purposes of dark matter, lies within a cold regime.

The solution of the SP system is carried out using our code CAFE-FDM described in \cite{Alvarez_Rios_2022}, with the following combination of numerical methods. The evolution uses the Method of Lines with a fourth order Runge-Kutta integrator, spatial discrete Fourier transform discretization of the right hand side of Sch\"odinger equation, and periodic boundary conditions on $\Psi$; the Poisson equation is solved using a Fast-Fourier-Transform based method over each intermediate step of the Runge-Kutta integrator. The evolution time-step uses the Courant-type condition $\Delta t / \Delta x^2 < \frac{1}{6\pi}$, as recommended in \cite{Chen2021}, that guarantees stability.

The evolution of these initial data using the SP system (\ref{eq:GP})-(\ref{eq:Poisson}) leads to virialized configurations that approach a virialized core-halo structure \cite{Chen2021,Curicavery2024}. The central region forms a solitonic core whose density profile, when averaged in space and time, closely matches the ground-state solution of the SP system obtained under isolated boundary conditions \cite{GuzmanUrena2004}. Surrounding this core, the outer halo is typically approximated with a Navarro-Frenk-White (NFW) profile, from structure formation simulations \cite{Schive:2014dra,Mocz:2017wlg}, as well as in local collapse   \cite{periodicas,PhysRevD.110.063502}.

In order to monitor the virialization, we compute the kinetic and potential energy observables inside a sphere of radius $r_t$, where $r_t$ denotes the transition radius between the solitonic core and the outer halo:

\begin{eqnarray}
    K_t &=& -\frac12 \int_{R\le r_t} \Psi^* \nabla^2 \Psi \, d^3x, \\
    W_t &=& \frac12 \int_{R\le r_t} V |\Psi|^2 \, d^3x.
\end{eqnarray}

\begin{figure}[H]
    \centering
    \includegraphics[width=8cm]{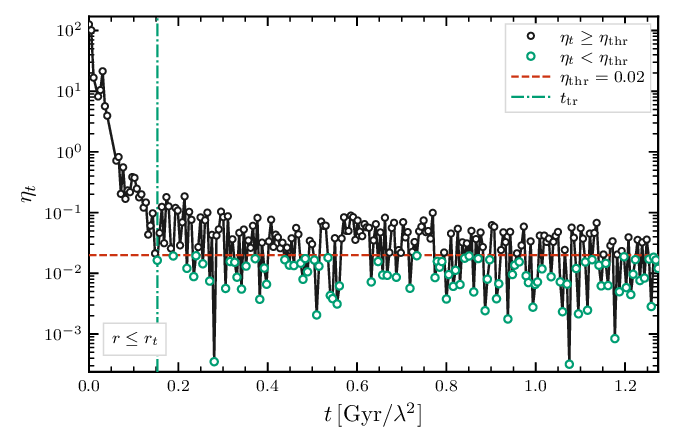}
    \includegraphics[width=8cm]{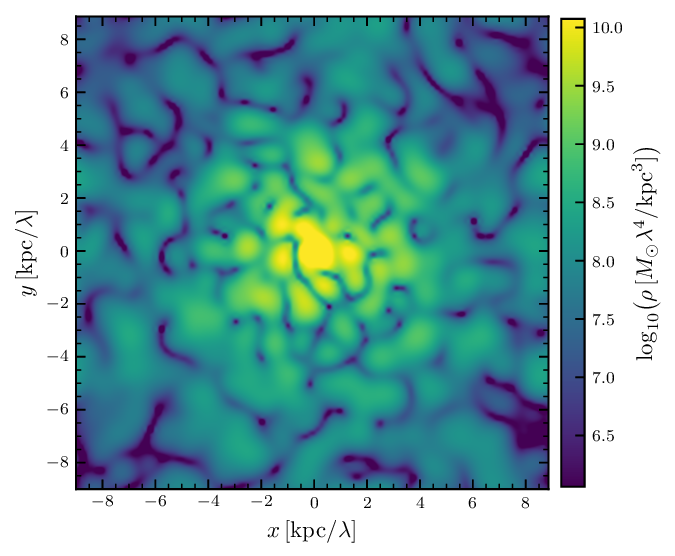}
    \caption{Virialization diagnostic and density structure at the transition time. (Top) evolution of the virial deviation parameter $\eta_t=|2K_t+W_t|/|W_t|$ calculated inside the transition radius $r_t$. The horizontal dashed line marks the adopted threshold $\eta_{\rm thr}=0.02$, and the vertical dash-dotted line indicates the first crossing time $t_{\rm tr}\simeq 0.153\,{\rm Gyr}/\lambda^2$. (Bottom) density on the $xy$ plane at $t_{\rm tr}$.}
    \label{fig:virial}
\end{figure}

\noindent For a virialized configuration, ideally these quantities combine to satisfy the relation $2K_t+W_t=0$. We therefore define

\begin{equation}
    \eta_t := \frac{|2K_t+W_t|}{|W_t|},    
\end{equation}

\noindent as the relative deviation from the ideal virial condition, normalized by the magnitude of the potential energy inside a sphere of radius $r_t$. The central structure is considered virialized when $\eta_t$ falls below a chosen tolerance threshold $\eta_{\rm thr}$, which is set here to $\eta_{\rm thr}=0.02$. The top panel of Fig.~\ref{fig:virial} shows the evolution of $\eta_t$ and identifies the first time at which $\eta_t<\eta_{\rm thr}$. This defines the transition time $t_{\rm tr}\simeq 0.153\,{\rm Gyr}/\lambda^2$. For later times, $\eta_t$ remains close to the threshold and oscillates around it. This diagnostic quantity shows that $t_{\rm tr}$ marks the beginning of a relaxed core-halo stage and provides a suitable reference time for the subsequent analysis. The bottom panel of Fig. \ref{fig:virial} shows the density distribution on the $xy$ plane at time $t_{\rm tr}$, where a condensed central core is already present. The core is not a smooth, spherically symmetric distribution, but a region where the density is high in angular average, whereas the surroundings composing the halo are lower density, however dominated by smaller granules, with higher kinetic energy and actually -although not seen in a snapshot- evolve in time moving around.

\subsection{Centering the core for analysis}

The result of the workhorse simulation defined above, is the space-time dependent complex wave function $\Psi$, from which the density is obtained as $\rho = |\Psi|^2$. A first step in the core-halo fitting is the identification of the halo center. Rather than using a simple maximum or center-of-mass over the full numerical domain, we construct a density-weighted centroid. In practice, this is implemented by selecting grid points above a given density threshold and calculating

\begin{equation}
\mathbf{x}_c = \frac{\sum \rho(\mathbf{x})\,\mathbf{x}}{\sum \rho(\mathbf{x})}.\label{eq:CM}
\end{equation}

\noindent This method provides a useful estimate of the core center. All radius dependent quantities are then calculated with respect to this center.

\subsection{Spherical Averaging and Denisty Model}

To separate large-scale structure from fluctuations, we construct a spherically averaged density profile. This is done by binning the density field in radial shells, where a solid angle average is calculated as

\begin{equation}
\rho_{{\rm model}}(r) = \langle \rho \rangle_{\Omega} = \dfrac{1}{4\pi} \int_{\Omega} \rho \, d\Omega,
\label{eq:rhoaveragetot}
\end{equation}

\noindent at a given time during the evolution, taken over all grid points within a shell centered at radius $r$ measured from the center location Eq. (\ref{eq:CM}), where $\Omega:=[0,\pi]\times[0,2\pi]$ denotes the solid angle over which the averaging is performed. The resulting angularly averaged density profile defines a radial density and a gravitational potential $\langle V \rangle$ through the Poisson equation (\ref{eq:Poisson}) that depends only on $r$.

Once the structure has collapsed and virialized, this averaged density profile is characterized by a central solitonic core surrounded by an outer region that decays like an NFW-like halo (see e.g. \cite{PhysRevD.110.063502}). In the bottom panel of Fig.~\ref{fig:virial} we show a slice of a snapshot of the already virialized workhorse structure with this core-halo distribution.

In the angular average, the solitonic core is modeled using the radius dependent profile \cite{Schive:2014dra}

\begin{equation}
\rho_{\text{core}}(r) = \rho_c \left[ 1 + 0.091\left(\frac{r}{r_c}\right)^2\right]^{-8},
\label{eq:soliton}
\end{equation}

\noindent that mimics the ground state stationary solution of the SP system \cite{GuzmanUrena2004}. This density model was proposed in \cite{Schive:2014dra}, as the result of structure formation simulations. In this formula, $\rho_c$ is the central density of the core whereas $r_c$ is the core radius, the radius at which the density is half the peak density. On the other hand, the outer region defined after a transition radius $r_t$, that is, for $r > r_t$, the distribution can be described in solid-angle average by the NFW profile

\begin{equation}
\rho_{\text{NFW}}(r) = \frac{\rho_s}{\frac{r}{r_s}\left(1 + \frac{r}{r_s}\right)^2}.
\label{eq:NFW}
\end{equation}

\noindent Adding these two components, the resulting angularly averaged density can be described by profile $\rho_{\text{model}}(r) = \rho_{\text{core}}(r)\Theta(r_t - r) + \rho_{\text{NFW}}(r)\Theta(r - r_t)$, where $\Theta$ is the Heaviside step function and $r_t$ denotes the transition radius between the core and the halo.

\begin{figure}[H]
\centering
 \includegraphics[width=\columnwidth]{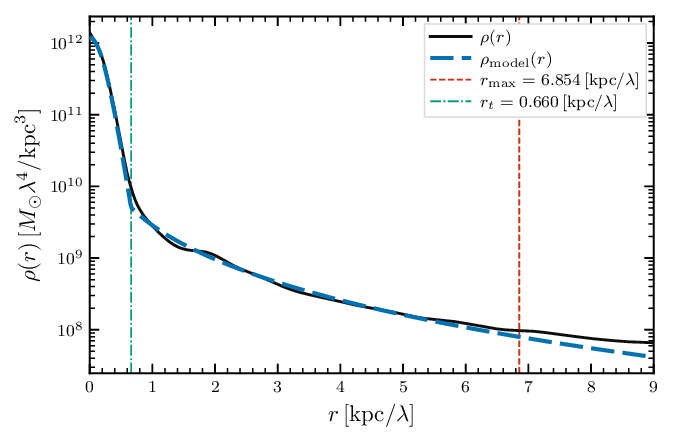}
 \includegraphics[width=\columnwidth]{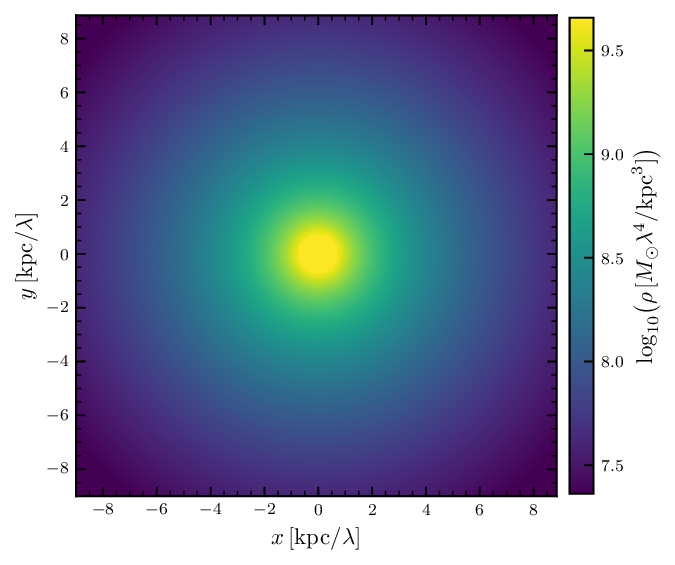} 
\caption{(Top) Density profile along $r$ at time $t_{\rm tr}$ with the values of the fitting parameters 
$r_s =3.529 ~ {\rm kpc}/\lambda$, $r_t = 0.660 ~ {\rm kpc}/\lambda$, 
$\rho_c = 1.31\times 10^{12} {\rm M}_\odot \lambda^4 / {\rm kpc}^3$ and 
$\rho_s = 1.03\times 10^{9}  {\rm M}_\odot \lambda^4 / {\rm kpc}^3$. 
(Bottom) Model density $\rho_{\rm model}(r)$ mapped back to the Cartesian domain.}
\label{fig:crude_modeldensity}
\end{figure}

In this way, $\rho_{\rm model}(r)$ captures the global core-halo structure, including the solitonic inner region and the extended halo. The profile is then interpolated back onto the full grid to construct a three-dimensional density $\rho_{\rm model}(\mathbf{x}) = \rho_{\rm model}(r(\mathbf{x}))$. This technical step is important since it defines a smooth background that can be removed if desired in order to study the fluctuations.

In the top panel of Fig.~\ref{fig:crude_modeldensity} we show the radial density profile along $r$, where the transition radius $r_t$ between the core and the NFW profile is indicated, as well as the radius $r_{\max}$ at which we perform an accurate fitting, trying to avoid the effect of the periodic boundary, which can be noticed from this radius on. At the same figure we show the smooth version $\rho_{\rm model}(r)$ embedded back into the 3D domain. The quality of the fit is evaluated over the fitting region $0<r\leq r_{\max}$.

The value $r_{\max}$ is estimated from the smoothed radial profile by identifying the point where the outer profile begins to flatten, which indicates that the boundary effects start to become relevant. The parameters of the model are then obtained through a least squares fit over the interval of confidence, leaving three free parameters, the core radius $r_c$, NFW scale $r_s$ and transition radius $r_t$, at time $t_{\rm tr}$ the best fit values are 
$r_s =3.529 ~ {\rm kpc}/\lambda$, $r_t = 0.660 ~ {\rm kpc}/\lambda$,
$\rho_c = 1.31\times 10^{12} {\rm M}_\odot \lambda^4 / {\rm kpc}^3$ and 
$\rho_s = 1.03\times 10^{9}  {\rm M}_\odot \lambda^4 / {\rm kpc}^3$. The fitted core-halo profile $\rho_{\rm model}$ uses a least squares algorithm, and  reproduces the density profile with a normalized RMS deviation of $\sim 5\%$, which improves when a smaller $r_{\rm max}$ is used.

It is worth mentioning that, even after virialization, the structure continues to evolve, never reaches a stationary state, and the granules in the halo never smooth out. This implies that granule positions and sizes evolve in time, therefore the fitted profile $\rho_{\rm model}$ also changes in time, and structures must be analyzed at various time slices to capture an average behavior of the structure in time as well.

\subsection{Residual density and spectral analisys}

We define the residual field as the difference between the full density and the spherical density model

\begin{equation}
\rho_{\rm res}(\bf x)= \rho(\mathbf{x}) - \rho_{\rm model}( {\bf x} ), 
\end{equation}

\noindent which isolates the fluctuations associated with interference patterns developed during the evolution.

Using a Fast Fourier Transform  ${\cal F}$, to
$\rho$, $\rho_{\rm model}$ and $\rho_{\rm res}$, we calculate the power spectrum defined as 

\begin{equation}
P(\mathbf{k}) = \left|{\cal F} \left[ \frac{\alpha - \bar{\alpha}}{\bar{\alpha}} \right] \right|^2,
\end{equation}

\noindent where $\alpha$ represents the three densities $\rho,~\rho_{\rm model},~\rho_{\rm res}$, and $\bar{\alpha}$ is the average of $\alpha$ in the numerical domain. We then calculate the dimensionless isotropic spectrum of these densities  in spherical shells in Fourier space

\begin{equation}
\Delta^2(k) = \frac{k^3}{2\pi^2} P(|\mathbf{k}|).
\end{equation}

\begin{figure}[H]
\centering
 \includegraphics[width=8cm]{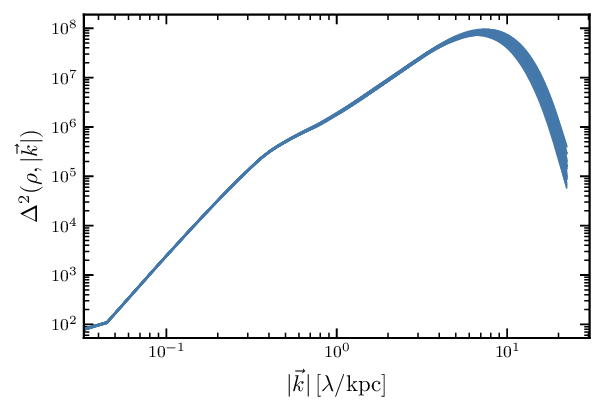}
 \includegraphics[width=8cm]{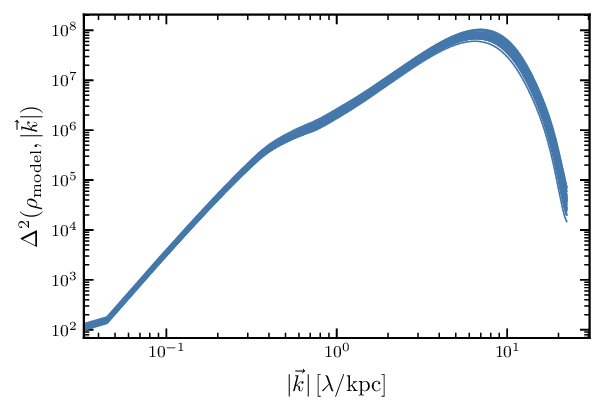}
 \includegraphics[width=8cm]{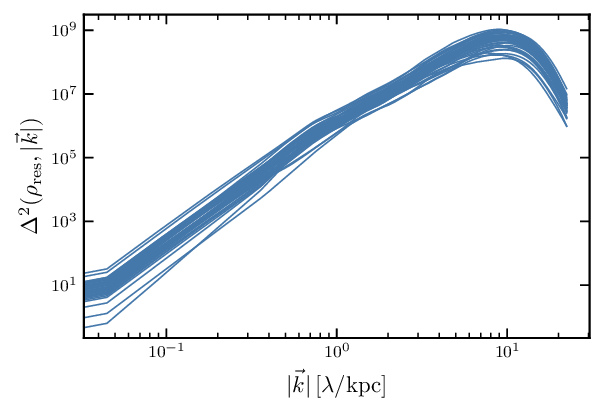}
\caption{Dimensionless power spectrum $\Delta^2(k)$ computed for $\rho$, $\rho_{\rm model}$ and $\rho_{\rm res}$. Each line corresponds to a different snapshot during the evolution in a lapse time when the configuration is already virialized. Notice that the evolution actually produces a time dependent spectrum more noticeable in $\rho_{\rm res}$, associated to the granular structure. }
\label{fig:spectrum}
\end{figure}

\noindent The resulting spectra are shown in Fig. \ref{fig:spectrum} for the total density $\rho$, the model density $\rho_{\rm model}$ and the granules $\rho_{\rm res}$ for 30 snapshots taken after the structure is virialized, between $t_{\rm tr}$ and the final time. First notice that the dynamics of the FDM changes with time for all three densities, including the model density.  However that the granular envelope has a more variable spectrum since it is more dynamic due to the dominant kinetic energy contribution in that region.

\begin{figure}[H]
\centering
 \includegraphics[width=8cm]{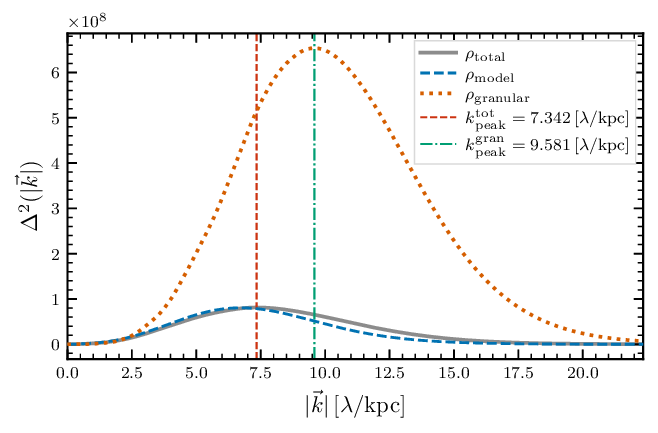}
\caption{Time average across the snapshots, of the dimensionless power spectra $\Delta^2(k)$ of the densities $\rho$, $\rho_{\rm model}$ and $\rho_{\rm res}$. Unlike the traditional log-scale spectrum, this one allows one to clearly observe the peak separation between peak frequencies. }
\label{fig:spectrumb}
\end{figure}

In order to clearly distinguish a difference between the peak frequencies of the whole structure with respect to that of the granular fluctuations, we show in Fig. \ref{fig:spectrumb} a linear scale plot of the mean of the spectra. For the density $\rho$, as well as for the model density $\rho_{\rm model}$, there are similar peaks at $|\vec{k}|\sim 0.7$, whereas for the granules this dominant peak appears at $|\vec{k}|\sim0.95$, which indicates preferred scales in the three fluctuation fields. This confirms the expected result that small fluctuations prefer a higher $|\vec{k}|$, and our analysis quantifies the difference between the two peaks. 

\section{Estimating the de Broglie wavelength}
\label{sec:dB}

The granular structure already implies that the velocity field across the structure has different directions and magnitudes. This suggests that, since $\lambda_{\rm dB}$ is velocity dependent, then its value depends on the region of the core-halo being considered. In order to estimate $\lambda_{\rm dB}$, we evaluate the density current 

\begin{equation}
j_k = \frac{1}{2i}\left(\Psi^* \partial_k \Psi - \Psi \partial_k \Psi^* \right),\nonumber
\end{equation}

\noindent where $k=x,y,z$, then estimate the magnitude of the velocity at each grid point of the numerical domain as 

\begin{equation}
{\bf v} = \frac{ {\bf j} }{\rho}, ~~~~~
v = \frac{\sqrt{|j_x|^2 + |j_y|^2 + |j_z|^2}}{\rho},\label{eq:velocity}
\end{equation}

\noindent which finally is used to estimate a position dependent de Broglie wavelength 

\begin{equation}
\lambda_{\mathrm{dB}}(r) = \frac{2\pi}{m_b \sigma_v(r)}.
\end{equation}

\begin{figure}[H]
\centering
 \includegraphics[width=8cm]{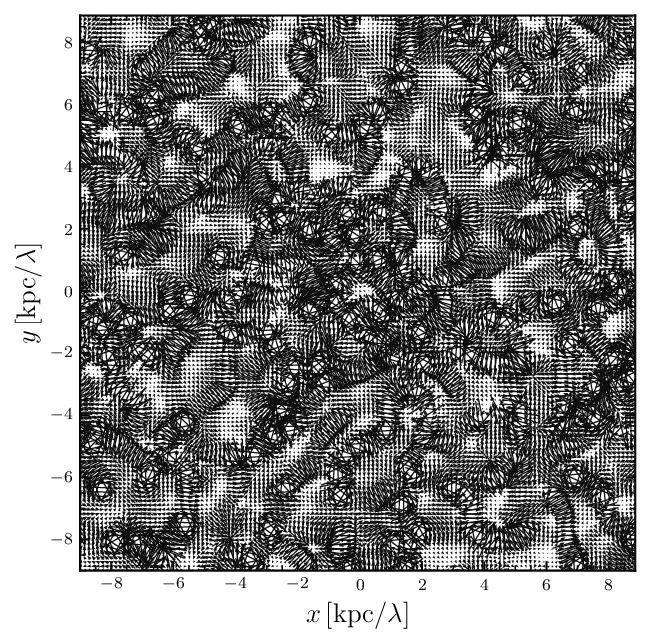}
\caption{Slice of the velocity field on the $xy-$plane for $m_B=10^{-23}$eV at time $t_{\rm tr}$, that illustrates how anisotropic the velocity field is.}
\label{fig:velocty}
\end{figure}

\noindent Here $\sigma_v(r)$ denotes the velocity dispersion computed over spherical shells at fixed radius

\begin{equation}
\sigma_v(r) = \left(\left\langle v^2 \right\rangle_{\Omega} - \left\langle v \right\rangle_{\Omega}^{2}\right)^{1/2},
\end{equation}

\noindent where $\langle\cdot\rangle_{\Omega}$ represents the angular average. In order to have an idea of how the velocity field (\ref{eq:velocity}) looks like in the snapshot we have been working with, in Fig. \ref{fig:velocty} we show the case for $m_B=10^{-23}$eV.

\begin{figure}[H]
\centering
 \includegraphics[width=\columnwidth]{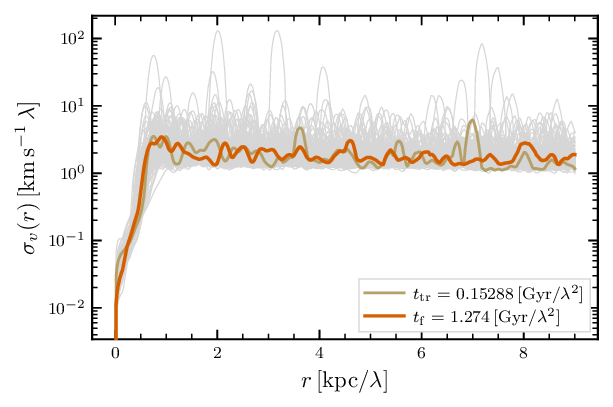}
 \includegraphics[width=\columnwidth]{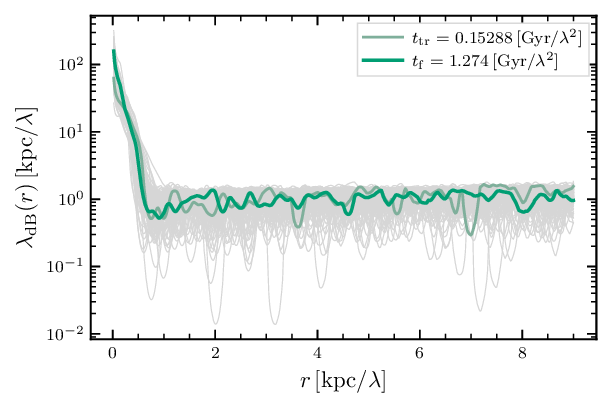}
 \caption{Radial profile of the velocity dispersion $\sigma_v(r)$ and the associated de Broglie wavelength $\lambda_{\rm dB}(r)$ from $t_{\rm tr}$ to $t_f$. Gray lines correspond to intermediate snapshots, while colored lines highlight $t_{\rm tr}$ and $t_f$.}
 \label{fig:dB}
\end{figure}

Figure~\ref{fig:dB} shows the radial dependence of the velocity dispersion $\sigma_v(r)$ and the corresponding de Broglie wavelength $\lambda_{\rm dB}(r)$ from the transition time $t_{\rm tr}$ to the final time of the simulation $t_f$. Since $\lambda_{\rm dB}$ is inversely proportional to $\sigma_v$, regions with smaller velocity dispersion correspond to larger de Broglie wavelengths. Inside the solitonic core, $\lambda_{\rm dB}$ is various orders of magnitude larger than in the outer halo, indicating that the core is dominated by long wavelength modes.

The radial profile of $\lambda_{\rm dB}(r)$ exhibits a nontrivial behavior, it decreases from the center, reaches a minimum at an intermediate radius, and increases again in the outer halo. Explicitly, $\lambda_{\rm dB}$ is maximum at the center of the configuration, within the core, as expected, since it is the region where the FDM is colder than the halo, since the kinetic energy and thus the velocity are smaller there than in the halo. At large radii, where the granules carry high kinetic energy, the velocity is expected to be higher and therefore $\lambda_{\rm dB}$ should be smaller. This result allows one to {\bf quantify} the difference in wavelength between the core and the halo. For the case of $m_{B}=10^{-23}$eV and the initial conditions used in the momentum space for our workhorse structure, we find that the minimum $\lambda_{\rm dB}$ at the interface is of $0.338~{\rm kpc}/\lambda$, whereas an average $\lambda_{\rm dB}$ from the transition region to the outskirts of the halo and averaged also in time is $\langle\lambda_{\rm dB}\rangle=1.033 ~{\rm kpc}/\lambda$.

The dominant peak of the granular spectrum provides a characteristic length scale, $2\pi/k_{\rm peak}^{\rm gran}$, which can be directly compared with the de Broglie wavelength. During the relaxed stage, we find $\langle\lambda_{\rm dB}\rangle=1.033,{\rm kpc}/\lambda$, while the mean granular size scale is $0.753,{\rm kpc}/\lambda$. Their ratio has a mean value of $0.714$. This shows that the characteristic scale extracted from the granular spectrum is systematically smaller than, but comparable to, the de Broglie wavelength. Therefore, the spectral peak scale and the de Broglie wavelength are connected at the same order of magnitude, supporting the interpretation that the observed granules are associated with de Broglie scale interference.

What is unexpected, is that there is a minimum precisely at the transition region between the core and the halo. The straightforward interpretation is that there the velocity should be higher than in the center and in the outer kinetic envelope. This is intriguing, although we find a precedent for the need of a high velocity at the transition sphere in the construction of core-halo FDM configurations within the Madelung frame \cite{AlvarezGuzmanMadelung}.

\section{Discussion and Conclusions}
\label{sec:conclusions}

We have presented a quantitative analysis of the small-scale structure of FDM halos, focusing on both the spectral properties of density fluctuations and the local velocity field.

From the decomposition of the density into a smooth core-halo component and a residual field, we showed that the interference pattern can be isolated and characterized through its power spectrum. The residual spectrum shows a well-defined peak that is clearly separated from the peak of the total and model densities, which identifies the characteristic scale of granules. 

Independently, we estimated a local de Broglie wavelength from the velocity field derived from the wavefunction. The resulting radial profiles reveal that $\lambda_{\mathrm{dB}}$ is strongly position-dependent and evolves in time, but consistently shows a non-monotonic behavior within the core and halo. In particular, we find that the wavelength reaches a minimum at an intermediate radius, close to the transition between the solitonic core and the outer halo.

These results add to the picture of the internal structure of FDM halos, where the core-halo boundary plays a central dynamical role. The identification of a preferred scale for fluctuations, together with its radial dependence, opens the possibility of connecting the granular structure to observable effects in real galaxies, such as dynamical heating or perturbations of stellar tracers.

Future work should extend this analysis to different halo masses, environments, and initial conditions, as well as explore the impact of these fluctuations on baryonic matter.

\section*{Acknowledgments}
This research is supported by
SECIHTI Grant No. CFB-2025-I-759,
Laboratorio Nacional de C\'omputo de Alto Desempe\~no Grant No. 2026-8, and CIC-UMSNH Grant No. 4.9.

\end{multicols}

\medline
\begin{multicols}{2}
\bibliographystyle{rmf-style}
\bibliography{ref}

\end{multicols}
\end{document}